\documentclass[preprint]{revtex4-2}

\usepackage{graphicx}
\usepackage{dcolumn}
\usepackage{bm}
\usepackage{physics}
\usepackage{amsmath,amssymb,mathtools}

\usepackage[utf8]{inputenc}
\usepackage[T1]{fontenc}
\usepackage{mathptmx}
\usepackage{etoolbox}

\begin{document}

\title{Nonexistence of Integrable Nonlinear Magnetic Fields with Invariants Quadratic in Momenta}

\author{B. Erd\'elyi}
 \affiliation{Department of Physics, Northern Illinois University, DeKalb, IL 60115}
 
\author{K. Hamilton}
\affiliation{Department of Physics, Northern Illinois University, DeKalb, IL 60115}

\author{J. Pratscher}
\affiliation{Department of Mathematics and Statistics, Stephen F. Austin State University, Nacogdoches, TX, 75962
}

\author{M. Swartz}
\affiliation{Department of Mathematics, Northern Illinois University, DeKalb, IL, 60115}

\date{\today}

\begin{abstract}
Nonlinear, completely integrable Hamiltonian systems that serve as blueprints for novel particle accelerators at the intensity frontier are promising avenues for research, as Fermilab's Integrable Optics Test Accelerator (IOTA) example clearly illustrates. Here, we show that only very limited generalizations are possible when no approximations in the underlying Hamiltonian or Maxwell equations are allowed, as was the case for IOTA. Specifically, no such systems exist with invariants quadratic in the momenta, precluding straightforward generalization of the Courant-Snyder theory of linear integrable systems in beam physics. We also conjecture that no such systems exist with invariants of higher degree in the momenta. This leaves solenoidal magnetic fields, including their nonlinear fringe fields, as the only completely integrable static magnetic fields, albeit with invariants that are linear in the momenta. The difficulties come from enforcing Maxwell equations; without constraints, we show that there are many solutions. In particular, we discover a previously unknown large family of integrable Hamiltonians.
\end{abstract}

\maketitle

\section{Introduction}

Particle accelerators constitute some of the most complex Hamiltonian systems
of practical importance. Traditionally, designs called for making them
as linear as possible. Nonlinearities were present as unwanted annoyances, or sometimes as single particle correction schemes. It was realized recently that to make further progress in beam dynamics, especially related to high intensity, space charge
dominated beams, it is necessary to make the systems themselves nonlinear on purpose in
order to alleviate problems with collective instabilities \cite{chao1993}. However, random
or regular multipolar nonlinearities have detrimental effects, such as
decreasing the region of orbital stability, the so-called dynamic aperture \cite{hermanDA}. An
option out of this conundrum is the potential existence of some "magic"
nonlinearities that make the underlying Hamiltonian system integrable in the
sense of Liouville \cite{arnold}. The main idea is to provide nonlinearities of significant magnitude in order to damp deleterious collective effects while not harming orbital stability and hence guaranteeing sufficiently large dynamic aperture. While the idea is not new \cite{McMillan}, a breakthrough recently paved the way towards the first practical nonlinear integrable particle accelerator \cite{bib:Danilov}. Indeed, these ideas have been successfully implemented in the Integrable Optics Test Accelerator (IOTA) \cite{Antipov_2017}.

In this paper, we investigate completely integrable Hamiltonian systems and associated magnetic fields beyond IOTA. The reasons are twofold: \textit{i)} to perform a comprehensive search for four-dimensional integrable Hamiltonians, \textit{ii)} without any approximations in the Hamiltonian itself and the Laplace equation satisfied by the magnetic fields. It is shown that integrable static nonlinear magnetic fields are scarce; in fact, the fringe fields of solenoids are the only such physically realizable fields. The associated invariants are linear in the momenta. Thus, unfortunately, we established a significant negative result: nonlinear magnetic fields that would make the Hamiltonian of a charged particle in external fields completely integrable, with invariants quadratic in momenta, do not exist. As a consequence, it is impossible to generalize the linear (Courant-Snyder) and IOTA (Danilov-Nagaitsev) results to the fully nonlinear case with no approximations.  We also conjecture that no such fields exist with invariants of higher degree in momenta. Overall, we are left with only invariants linear in momenta, which are straightforward consequences of Noether's theorem \cite{Noether1918}. These results deepen our understanding of nonlinear integrable optics accelerators and point the way towards alternate avenues for advancing the intensity frontier in beam physics.

Furthermore, our search for nonlinear integrable Hamiltonians of the same form, but not physically constrained, turned out more positively. On one hand, this is hardly surprising since the topic has a long-standing interest in dynamical systems, and the relevant literature consists of many possible solutions for various cases \cite{bib:Dorizzi,McSween,Berube,Pucacco2004,Kubu2023,Heinzl2017,Hoque2024}. However, one major difference in the definition of magnetic fields should be pointed out: for us, the magnetic fields are so-called external fields, that is solutions of the homogeneous Maxwell equations in the absence of any electric fields. These are the physically realizable static magnetic fields in a source-free region of space. These magnetic fields are both divergence- and curl-free, therefore being derivable from a "magnetic scalar potential" that satisfies the Laplace equation. On the other hand, virtually all papers in this area consider "magnetic fields" as being derivable from a vector potential, compatible with the fields being divergence-free, but not imposing the vanishing curl. As we show, this has tremendous consequences to the size of the solution space. If one would argue that the curl could be non-vanishing in the presence of some current density, that would not be compatible with the stated form of the Hamiltonian, i.e. a single charged particle in prescribed external fields. 

 We studied a special case without the constraints imposed by the vanishing curl of the "magnetic fields". The culprit in the scarcity of results alluded to in the preceding paragraph is mainly due to Maxwell equations. Indeed, we show that not imposing any physically relevant constraint leads to many interesting solutions. We delve more deeply into a large family of Hamiltonians with trigonometric "magnetic fields" in Section V. These results add to our fundamental understanding of integrability from a mathematical perspective, but we barely scratched the surface of this important topic here, and much work remains to be done.

The full paper is organized as follows: in Sections II and III, we establish the foundations of the search for integrable Hamiltonian systems with no approximations and determine the general form of the invariants.  Section IV discusses the case of the second invariant being linear in momentum and provides an avenue to show that the solenoidal fields satisfy the requirements for this integrable system.  Section V then considers an invariant quadratic in momenta, and is broken up into four subsections according to the symmetry-based interpretation of the invariant. In addition, there is a diagonal Cartesian subsection for non-Maxwellian magnetic fields. Finally, Section VI describes the generalization of the theory to higher orders in momenta.

\section{Problem Setup}

The Hamiltonian of a classical point-like charged particle of mass $m$ and charge $q$ in external electromagnetic fields is well known \cite{bib:Jackson}:
\begin{equation}
    H = c\sqrt{(\mathbf{p}-q\mathbf{A}(\mathbf{x},t))^{2}+m^{2}c^{2}}+q\Phi(\mathbf{x},t).
    \label{CEDHamiltonian}
\end{equation}
However, in accelerator physics, the independent variable is usually the
arclength along the trajectory of a reference particle, and the canonical
coordinates are defined relative to this fiducial \cite{berz1999modern}. To this end, the Hamiltonian can be
expressed as \cite{bib:Wolski}%
\begin{equation}
    H = \frac{\delta}{\beta_{0}}-\sqrt{\qty(\delta+\frac{1}{\beta_{0}}-\phi)^{2}-(a-a_x)^2
    -(b-a_y)^2-\frac{1}{\gamma_{0}^{2}\beta_{0}^{2}}}-a_{s},
    \label{eq:fullH}%
\end{equation}
where $\left(  x,a,y,b,\delta\right)  $ are canonical coordinates, $s$ is the
arclength that plays the role of time, and $\left(  a_{x},a_{y},a_{s}\right)
$ are the Cartesian components of the scaled vector potential. While $\left(
x,y\right)  $ are the usual spatial coordinates, $\left(  a,b\right)  $ are
scaled canonical momenta in the $x$ and $y$ directions, respectively. The
scaling factor is $q/p_{0}$ for the vector potential, $q/(cp_0)$ for the scalar potential, and $1/p_{0}$ for the momenta, where $p_{0}$ is the momentum of the
reference particle. Finally, $\delta$ is the relative total energy (kinetic +
potential) deviation with respect to the reference particle. The subscript $0$ refers to the reference particle; $\beta_0, \gamma_0$ being the usual relativistic quantities.

To study the integrability of this Hamiltonian seems like a formidable task. However, we can
simplify the problem by first noticing that for static magnetic fields, $\phi$ vanishes and $\delta$
is a constant, so it becomes a simple parameter in (\ref{eq:fullH}). Second,
we can use gauge freedom to our advantage. That is, we choose a gauge where
$a_{s}$ vanishes (axial gauge). A simple constructive proof shows that this is always
possible, and the gauge invariant (scaled) magnetic field components $(b_x, b_y, b_s)$ relate to the vector potential by%
\begin{align}
a_{x}  & =\int_{s_{0}}^{s}b_{y}\left(  x,y,t\right)  dt,\\
a_{y}  & =-\int_{s_{0}}^{s}b_{x}\left(  x,y,t\right)  dt+\int_{0}^{x}%
b\,_{s}\left(  t,y,s_{0}\right)  dt,\\
a_{s}  & =0.
\end{align}
The arbitrary reference point $s_{0}$ is typically chosen to be zero. Under
these circumstances, it is easy to see that if $I$ is an autonomous constant of the motion
for $H,$ $dI/ds=[H,I]=0$, then it is also a first integral for%
\begin{equation}
H_{e}\left(  x,a,y,b;s\right)  =\left(  a-a_{x}\left(  x,y,s\right)  \right)
^{2}+\left(  b-a_{y}\left(  x,y,s\right)  \right)  ^{2},
\label{eq:eqH}%
\end{equation}
since the square root can never vanish. Therefore, we reduced the study of
integrability of (\ref{eq:fullH}) to the two degree of freedom
``time-dependent" Hamiltonian integrability study of (\ref{eq:eqH}).

The integrability of this Hamiltonian has been studied before in the
autonomous setting, first in \cite{bib:Dorizzi}, followed by many other more recent works \cite{McSween,Berube,Pucacco2004,Kubu2023,Hoque2024}. Although no complete solution set has been found to date, there are several special cases of interest. While time dependence of the Hamiltonian is not a
stringent drawback, since by extending the phase space it can be made
autonomous, it does require an extra independent integral in involution for
complete integrability. The other avenue is a direct search for first integrals in the time dependent case. Even if this is tried, the possible functional forms for
these invariants is huge. Based on earlier results from the literature, it is apparent that a classification of the integrals based on their momentum dependence is a
fruitful one. While there are some rare cases where the integrals are
transcendental functions of the momenta \cite{HietarintaTranscendental,GiacominiHigherTranscendental,GrammaticosLog}, it also appears that most of these
can be reformulated as invariants polynomial in momenta \cite{HallConvertPoly}. Furthermore, if
the polynomial forms are studied, again it is clear that invariants of high
order in momenta are rare \cite{NakagawaCompleteList,NakagawaNew}. Therefore, it seems that invariants of first and
second order in momenta have the best chance of success. This choice is also
the most important practically, as supported by the fact that in the conventional linear setting the well-known (Courant-Snyder) invariants are quadratic in momenta \cite{CourantSnyder}. The same is true for IOTA \cite{bib:Danilov}. Then, one could view the fully nonlinear setting without approximations in the Hamiltonian and the Laplacian as a canonical generalization.

We start by linearizing the vector potential; then the resulting quadratic Hamiltonian, which corresponds to a linear system, is always integrable \cite{KocakLinearIntegrability}. However, the normal form obtained by the standard linear time-dependent normalizing canonical transformation does not
leave the Hamiltonian form-invariant \cite{bib:Danilov}. Our goal is to obtain a Hamiltonian that
is time-independent and form-invariant. To this end, several coordinate
transformations are possible that leave the integrability of the Hamiltonian intact. Some of these are translations, scalings, and rotations \cite{HietarintaDirectSearch}. Indeed, it can be shown that the linearized system can be made $s$-independent with these transformations. In general, the transformations will be $s$-dependent.
In other words, we established equivalence classes of functions that serve as completely integrable Hamiltonians that map into each other under simple transformations that preserve integrability. Most importantly, the Hamiltonian becomes time independent, so it is a
conserved quantity. So, the strategy is to add a nonlinear part to the vector
potential in transformed coordinates, making the full Hamiltonian a constant
of motion, and if we can find an independent additional first integral such
that $\left[  H_{e},I\right]  =0$, we obtained complete integrability. Here
$\left[  \cdot,\cdot\right]  $ stands for the Poisson bracket in scaled
coordinates. If we are successful, we can prove the integrability of the
original system (\ref{eq:fullH}). So, we focus on the Hamiltonian in
\textquotedblleft normalized" coordinates obtained by translation, rotation and scaling to remove the $s$-dependence,%
\begin{equation}
H_{e,n}=\left(  a_{n}-a_{x}\left(  x_{n},y_{n}\right)  \right)  ^{2}+\left(
b_{n}-a_{y}\left(  x_{n},y_{n}\right)  \right)  ^{2}.
\label{eq:normalH}
\end{equation}
Heretofore, we drop the $_{n}$ for notational simplicity since there is no
danger of confusion.

In view of the preceding discussion, we search for invariants linear and quadratic
in momenta. More precisely, following Dorizzi \cite{bib:Dorizzi}, we look for first integrals linear and quadratic in the velocities.

\section{Problem Branching}

Consider the Hamiltonian
\begin{equation}
    H = \frac{1}{2}\qty(p_{x}^{2}+p_{y}^{2})+A(x,y)p_{x}+B(x,y)p_{y}+W(x,y),
    \label{DorizziHamiltonian}
\end{equation}
where $A$, $B$ and $W$ are arbitrary functions of their arguments.  Equation (\ref{DorizziHamiltonian}) may or may not be completely integrable based on the particular values of these functions.  However, even if the Hamiltonian turns out to be completely integrable for certain $A$, $B$, and $W$, this Hamiltonian must still obey Maxwell's equations in order to serve as a model for novel accelerators.

Setting $A$ and $B$ to zero, such that
\begin{equation}
    H = \frac{1}{2}\qty(p_{x}^{2}+p_{y}^{2})+W,
    \label{IotaH}
\end{equation}
we obtain Hamiltonians in ``natural" form, which leads to IOTA-type Hamiltonians (paraxial approximation) for $W=-a_s$. This has been extensively studied \cite{bib:Danilov}, and it was shown to have completely integrable elements, with nonlinear transverse magnetic fields that satisfy an approximation of the Laplace equation, in which the $s$-dependent derivatives are neglected.

On the other hand, if $A=a_x, B=a_y$, while $W=a_x^2+a_y^2$, we obtain the Hamiltonian (\ref{eq:normalH}). Furthermore, using the methods of Dorizzi \cite{bib:Dorizzi}, (\ref{DorizziHamiltonian}) can be put into a gauge invariant form by specifying $b_s$ and $U$ as
\begin{equation}
    b_s = \frac{\partial B}{\partial x}-\frac{\partial A}{\partial y},
    \label{WantedBField}
\end{equation}
and
\begin{equation}
    U = W - (A^2+B^2).
\end{equation}
Clearly, for our case $U=0$.  Maxwell's equations must govern the fields under consideration; as usual, the divergence of the magnetic field under consideration must vanish, allowing for the magnetic field to be found as the curl of a vector potential,
\begin{equation*}
    \nabla\cdot \mathbf{B} = 0 \quad \Rightarrow \quad \mathbf{B} = \nabla\cross\mathbf{A},
\end{equation*}
in agreement with \eqref{WantedBField}.  However, since we are also considering a source-free region, the curl of the magnetic field must also vanish,
\begin{equation*}
    \nabla\cross\mathbf{B} = 0.
\end{equation*}
This immediately allows us to obtain our magnetic field, $b_{s}$ from a magnetic scalar potential $V$ that satisfies the Laplace equation,
\begin{align}
    b_s &= \frac{\partial V}{\partial s}, \\
    \Delta V &= 0.
\end{align}
Also, following the gauge transformation, the Hamiltonian (\ref{DorizziHamiltonian}) reduces to
\begin{equation}
    H = \frac{1}{2}\qty(\dot{x}^{2}+\dot{y}^{2}).
    \label{eq:GenHamiltonian}
\end{equation}
In order for the system to be classified as completely integrable, there must be another invariant in involution with this Hamiltonian, which is taken to be of the general form
\begin{equation}
    I = g_{0}\dot{x}^{2}+g_{1}\dot{x}\dot{y}+g_{2}\dot{y}^{2}+k_{0}\dot{x}+k_{1}\dot{y}+h,
    \label{eq:QuadInv}
\end{equation}
where the functions $g_0, g_1, g_2, k_0, k_1, h$ are arbitrary functions of $x$ and $y$, subject to the constraint $\Delta b_s=0$. In the following sections we demonstrate that solutions exist only for $g_0=g_1=g_2=0$.

\section{Linear Invariants}
We now consider the velocity-dependent Hamiltonian of the form \eqref{eq:GenHamiltonian}. The linear invariant is
\begin{equation}
    I = k_{0}\dot{x}+k_{1}\dot{y}+h,
\end{equation}
so taking the time derivative and removing the factors of acceleration, we find the following system of equations:
\begin{align}
    k_{0,x} &= 0, \\
    k_{1,y} &= 0, \\
    k_{0,y}+k_{1,x} &= 0, \\
    h_{x} - k_{1}b_{s} &= 0, \label{LinearHx} \\
    h_{y} + k_{0}b_{s} &= 0, \label{LinearHy}
\end{align}
where $_{,x}$ and $_{,y}$ indicate partial derivatives with respect to $x$ and $y$. It is clear that $k_{0}$ is a function of $y$ only, and likewise $k_{1}$ is a function of $x$.  In combination with the third condition, we find that $k_{0}$ and $k_{1}$ become
\begin{equation}
\begin{split}
    k_{0} &= \alpha y+\beta, \\
    k_{1} &= -\alpha x+\gamma,
\end{split}
\label{LinearKFunctions}
\end{equation}
for some real constants $\alpha, \beta, \gamma$. We then substitute these two results back into the invariant,
\begin{equation}
    I = -\alpha\qty(x\dot{y}-y\dot{x})+\beta\dot{x}+\gamma\dot{y}+h.
\end{equation}
With the invariant of this form, we need to find those values of $\alpha$, $\beta$, and $\gamma$ such that this system is physically realizable.  We recognize from the compatibility condition $h_{xy}=h_{yx}$ that, if we take the appropriate derivative of \eqref{LinearHx} and \eqref{LinearHy}, we may set them equal to each other to find
\begin{equation}
    (-\alpha x+\gamma)b_{s,y}+(\alpha y+\beta)b_{s,x} = 0. \label{LinearCompatibility}
\end{equation}
This provides us with two possible cases to examine: $\alpha=0$ and $\alpha\neq 0$.  In the case of $\alpha=0$, we can integrate over $x$ and $y$, and we may write $b_{s}$ as a function $g^{\prime}$ (prime denoting derivative with respect to its argument)
\begin{equation}
    b_{s} = g^{\prime}(\gamma x-\beta y),
    \label{LinearTranslationalB}
\end{equation}
such that when substituting this back into \eqref{LinearHx} and \eqref{LinearHy}, we find
\begin{equation}
    h = g(\gamma x-\beta y),
\end{equation}
where, again, $g$ is an arbitrary function of its argument.  When we have $\alpha\neq 0$, we may use a scaling to set $\alpha=1$ and a translation in $x$ and $y$ to reduce \eqref{LinearCompatibility} to
\begin{equation*}
    yb_{s,x}-xb_{s,y} = 0.
\end{equation*}
In a similar style to the case of $\alpha=0$, we may then integrate over $x$ and $y$ to find a form for $b_{s}$ and $h$,
\begin{align}
    b_{s} &= \frac{-1}{\sqrt{x^{2}+y^{2}}}\,g^{\prime}(\sqrt{x^{2}+y^{2}}), \\
    h &= g(\sqrt{x^{2}+y^{2}}).
\end{align}

With these two subcases of $\alpha=0$ and $\alpha\neq 0$, we have completed the study of completely integrable Hamiltonians with second invariants linear in the momenta. This case boils down to the usual Noether’s theorem and hence it is related to the usual conservation principles in physics. Unsurprisingly, we found that those Hamiltonian systems are integrable in this class that have a magnetic field that is invariant with respect to a translation along an arbitrary line in the transverse plane. No such nonlinear magnetic field exists that satisfies the Laplace equation. The other class of integrable Hamiltonians have magnetic fields that are invariant with respect to arbitrary rotations in the transverse plane. Fortunately, there are such magnetic fields that satisfy the full Laplace equations: solenoids, including fringe fields.

To see this, consider the most general form of a scalar magnetic potential in a source free region in cylindrical coordinates \cite{ErdelyiBerzLindemann}:
\begin{equation}
V(r,\phi,s) = \sum_{l=0}\qty[f_{l}(r,s)\sin(l\phi)+g_{l}(r,s)\cos(l\phi)]r^{l},
\label{ScalarMagPot}
\end{equation}
where $r=\sqrt{x^{2}+y^{2}}$.  Taking a derivative with respect to $s$ gives us the $s$-component of the magnetic field, which we can write as
\begin{equation}
    b_{s} = \sum_{l=0}\qty[f_{l}^{\prime}(r,s)\sin(l\phi)+g_{l}^{\prime}(r,s)\cos(l\phi)]r^{l}.
\end{equation}
Also notice that this function can be written in terms of $x$ and $y$ through a coordinate transformation, thereby allowing us to write $b_{s}$ as
\begin{equation}
b_{s} = \sum_{k=0}\sum_{j=0}a_{k,j}(s)x^{k}y^{j}.
\label{LinearMagField}
\end{equation}
Meanwhile, we have seen the form of $b_{s}$ that results in translational invariance arises from \eqref{LinearTranslationalB}, which may also be written in terms of a power series
\begin{equation*}
b_{s} = \sum_{l=0}c_{l}(s)\qty(\gamma x-\beta y)^{l}.
\end{equation*}
However, this cannot have a vanishing Laplacian if $l\geq 2$, unless $\gamma=\beta=0$, thereby requiring
\begin{equation*}
    b_{s} = c_{0}+c_{1}\qty(\gamma x-\beta y).
\end{equation*}
Therefore, there are no nonlinear fields that are translational-invariant with respect to some direction in the $(x,y)$ plane with vanishing Laplacian.

On the other hand, \eqref{ScalarMagPot} will be invariant under arbitrary rotations when the expansion consists of only the $l=0$ terms.  In this case, the trigonometric functions evaluate to 0 and 1, respectively, and $V$ is written simply as
\begin{equation}
V = g_{0}(r,s).
\end{equation}
The function $g_{0}(r,s)$ can be found using a summation over the derivatives of the skew multipoles, $a_{l,l}(s)$,
\begin{equation*}
g_{l}(r,s) = \sum_{n=0}^{\infty}\frac{a_{l,l}^{(2n)}(s)}{\prod_{\nu=1}^{n}\qty(l^{2}-\qty(l+2\nu)^{2})}r^{2n},
\end{equation*}
where the exponent $(2n)$ denotes the derivative of order $2n$.
The magnetic field may then be found from the partial derivative of $V$ with respect to $s$,
\begin{equation}
b_{s} = a_{0,0}^{\prime}(s)-\frac{a_{0,0}^{(3)}(s)}{4}r^{2}+\frac{a_{0,0}^{(5)}(s)}{64}r^{4}+\dots.
\end{equation}
This is clearly a non-linear magnetic field that is invariant upon arbitrary rotations (but not translations) in the transverse plane.  Therefore, finite solenoids, including their fringe fields, do admit a second invariant linear in velocities. It is worth noting that our definition of solenoidal fields include finite length solenoids (the reason for the fields being nonlinear); it is sometimes customary to call solenoidal fields the field generated by only infinitely long solenoids, that is constant fields. 

\section{Quadratic Invariants}

As we have seen in the linear case, the existence of sufficiently many invariants in involution to make the Hamiltonian completely integrable is tied to symmetries. At first thought, this might not seem very surprising since the well-known Noether’s theorem states roughly that to every continuous symmetry of the system there is an associated conserved quantity. However, there is more to it: complete integrability needs sufficiently many invariants that are functions of momenta and coordinates, not all of which may be – in principle – associated with symmetries. But it turns out that this is indeed true in the quadratic case, namely higher order symmetries must exist to “generate” the needed invariants \cite{Perelomov1989Integrable}. This realization led to the setup of a Lie algebraic structure in the sense that the elementary symmetries form a basis for a Lie algebra, and the Lie group associated with this algebra acts on the Hamiltonian system and leaves it invariant. The orbits of the Lie group are associated with the invariants of the system.
The elements of the Lie algebra are the horizontal and vertical translations, and rotations; the corresponding elementary conserved quantities are the horizontal and vertical momenta, and the angular momentum. The conserved quantities generically can be written as a linear combination of these elements, where the coefficients are arbitrary functions of the coordinates (to be determined for integrability).

If we start with the general form for the invariant quadratic in the momenta (\ref{eq:QuadInv}),
and taking the total time derivative, as before, we find the following system of equations:
\begin{align}
g_{0,x} = 0, \quad & \quad g_{2,y} = 0, \\
g_{0,y}+g_{1,x} = 0, \quad & \quad g_{1,y}+g_{2,x} = 0, \\
k_{0,x}-g_{1}b_{s} = 0, \quad & \quad k_{1,y}+g_{1}b_{s} = 0, \\
h_{x}-k_{1}b_{s} = 0, \quad & \quad h_{y}+k_{0}b_{s} = 0, \\
k_{1,x}+k_{0,y} &= 2b_{s}\qty(g_{2}-g_{0}),
\end{align}
where $b_s$ is the $s$-component of the magnetic field and again must satisfy the Laplace equation.  The general solutions to these $g_{i}$'s can be found to be of the form
\begin{align}
    g_{0} &= ay^{2}-b y+c, \\
    g_{1} &= -2a xy+b x-d y+e, \\
    g_{2} &= a x^{2}+d x+f,
\end{align}
where $a,b,c,d,e,f$ are real numbers. Substituting these three functions back into the invariant, we can then rewrite it as
\begin{equation}
    I = a\qty(x\dot{y}-y\dot{x})^{2}+\qty(x\dot{y}-y\dot{x})\qty(b\dot{x}+d\dot{y})+c\dot{x}^{2}+f\dot{y}^{2}+e\dot{x}\dot{y}+k_{0}\dot{x}+k_{1}\dot{y}+h,
    \label{QuadInvariantGeneral}
\end{equation}
the form that makes the generators of the Lie algebra explicit (translations and rotations). It is with this form, then, that we can conduct transformations to the invariant, to include rotations, translations, and linear combinations with the Hamiltonian, such that the invariant is reduced to eight different cases based upon the geometry of the invariant and corresponding symmetries.  These cases are defined in Table \ref{Table1}, according to the values of the constants in the invariant of \eqref{QuadInvariantGeneral}.  Notice that the Cartesian, parabolic, and elliptical cases in general also have degenerate forms of the invariant, which arise from the conduction of rotations, such that they will include complex components.  However, we can safely discard the degenerate cases for the purpose of this paper. The reason is that the magnetic fields that we look for as solutions should be physically realizable, hence real-valued. This makes the Hamiltonian real-valued. In the degenerate cases, we can split the invariants into their real and imaginary parts. Then, both must be invariants by themselves. Hence, those cases would have value on their own only if the two invariants were independent, possibly resulting in super-integrable systems. However, it is easy to see that the real parts of both the degenerate and non-degenerate cases will always match. Since we are looking for real magnetic fields, it is expected that real solutions will not exist in the degenerate cases when solutions do not exist in the non-degenerate cases.  We therefore will exclude the degenerate cases from our search.
\begin{table}
\caption{List of each case of a quadratic invariant in the momentum defined by the constants in the general form \eqref{QuadInvariantGeneral}.}
    \centering
    \begin{tabular}{l r}
    \multicolumn{2}{c}{Invariant Cases (Orbits)} \\
    \hline\hline
    Case Geometry & Constant Values \\
    \hline
    Cartesian & $a=b=d=0$, $c\neq f$, $e/(c-f)\neq \pm i$ \\
    Degenerate Cartesian &  $a=b=d=0$, $c\neq f$, $e/(c-f)=\pm i$ \\
    Diagonal Cartesian & $a=b=d=0$, $c=f$ \\
    Parabolic & $a=0$, $b^{2}+d^{2}\neq0$ \\
    Degenerate Parabolic & $a=0$, $d = \pm ib\neq 0$ \\
    Spherical &  $a\neq 0$, $\lambda_{1}=\dfrac{bd}{2a}-e=0$, $\lambda_{2}=\dfrac{b^{2}-d^{2}}{4a}-c+f = 0$ \\
    Elliptical & $a\neq 0$, $\lambda_{1}\neq i\lambda_{2}\neq 0$ \\
    Degenerate Elliptical & $a\neq 0$, $\lambda_{1}=\pm i\lambda_{2}\neq 0$ \\
    \hline\hline
    \end{tabular}
    
    \label{Table1}
\end{table}

In order to conduct the search in each of the cases defined in Table \ref{Table1}, we recall that $b_{s}$ must satisfy the Laplace equation, hence it is harmonic, therefore it is real analytic inside the domain of interest to us, allowing us to write it as a power series,
\begin{equation}
b_{s} = \sum_{k=0}\sum_{j=0}a_{k,j}x^{k}y^{j}.
\label{PowerSeriesB}
\end{equation}
Furthermore, we may use the Taylor series representation where $a_{k,j}=\frac{b_{k,j}}{k!\,j!}$, where the coefficients are related as 
\begin{equation}
b_{k,j+2}=-b_{k+2,j}, \label{HarmonicCondition}
\end{equation} 
which we will refer to as the harmonic condition.  This condition will be used for all cases with the exception of the diagonal Cartesian case.  The Cartesian case's "normal form" still contains a free-parameter \cite{Winternitz}; in this sense, the diagonal Cartesian case is a non-degenerate Cartesian case with a specific value of the free parameter, equivalent to setting $c_1=0$ in (39) of \cite{Winternitz}. The suspension of the harmonic condition leads to interesting results in the diagonal Cartesian case, as we describe below.

\subsection{Diagonal Cartesian Case - Suspension of the Harmonic Condition}
This subsection summarizes our generic mathematical results, without the physical constraints. That is, we begin our investigation into each of the cases listed in Table \ref{Table1} with the diagonal Cartesian case and suspending the requirement of the harmonic condition, thereby allowing non-Maxwellian fields to be included as possible solutions. The Cartesian case has been considered previously in the literature and shown to admit solutions when arbitrary fields have been allowed in both a classical and quantum system in two and three dimensions \cite{bib:Dorizzi,Zhalij2015}. In 2D generic results exist, while in 3D only special cases. However, while the diagonal Cartesian case may be considered as the usual Cartesian case with fixed parameters, this case is the only one that allows for simultaneous elimination of the $\dot{x}^{2}$ and $\dot{y}^{2}$ terms in the invariant due to the form of the Hamiltonian.  This is done when $a=b=d=0$ while $c=f$, such that the invariant may be written as
\begin{equation}
    I = c\qty(\dot{x}^{2}+\dot{y}^{2})+e\dot{x}\dot{y}+k_{0}\dot{x}+k_{1}\dot{y}+h.
    \label{DiagCartInv1}
\end{equation}
We may reduce the form of the invariant through the use of integrability-preserving transformations, which will provide an equivalence class for the invariant in the general form.  One such transformation, which we use here, is to take a linear combination between the invariant \eqref{DiagCartInv1} and a multiple of the Hamiltonian, $\lambda H$, where $H$ is defined in \eqref{eq:GenHamiltonian}.  This linear combination will still enforce the necessary condition that the Poisson bracket and the new invariant function will vanish.  Choosing a multiplicative factor of $\lambda=-2c$ will then remove the first term from \eqref{DiagCartInv1}, and the invariant reduces to
\begin{equation}
    I = e\dot{x}\dot{y}+k_{0}\dot{x}+k_{1}\dot{y}+h.
    \label{DiagCartInv2}
\end{equation}
We may then scale $e\to 1$, which allows us to write a much simplified form of our system of equations,
\begin{align}
    g_{0} = 0, \quad & \quad g_{1} = 1, \label{DiagCartSysEq1} \\
    g_{2} = 0, \quad & \quad k_{0,x} = b_{s}, \\
    k_{1,y} = -b_{s},\quad & \quad k_{0,y}+k_{1,x} = 0, \label{DiagCartSysEq3} \\
    h_{x} = k_{1}b_{s},\quad & \quad h_{y} = -k_{0}b_{s}. \label{DiagCartSysEq4}
\end{align}
Since we are not restricting ourselves to considering Maxwellian fields, we will forgo the power series representation for $b_{s}$ for the time-being, allowing us to write the full form of the $k_{i}$ coefficients as
\begin{align*}
    k_{0} &= \int b_{s}\dd{x}+f(y), \\
    k_{1} &= -\int b_{s}\dd{y}-g(x).
\end{align*}
Taking the appropriate derivatives of of these expressions according to the second condition in \eqref{DiagCartSysEq3}, the linear combination between the two then gives
\begin{equation}
    \qty(\int b_{s}\dd{x})_{y}+f^{\prime}(y)-\qty(\int b_{s}\dd{y})_{x}-g^{\prime}(x) = 0.
    \label{DiagCartCond1}
\end{equation}
This equation then has solutions for the case of $f\equiv g\equiv 0$, and $b_{s}$ may be found as the product of two trigonometric or hyperbolic functions in $x$ and $y$.  The full set of possible basis function solutions for $b_{s}$ are
\begin{align*}
    b_{s,1} = \cos(kx)\cos(ky), \quad & \quad b_{s,2} = \cos(kx)\sin(ky), \\
    b_{s,3} = \sin(kx)\sin(ky), \quad & \quad b_{s,4} = \sin(kx)\cos(ky), \\
    b_{s,5} = \cosh(kx)\cosh(ky), \quad & \quad b_{s,6} = \cosh(kx)\sinh(ky), \\
    b_{s,7} = \sinh(kx)\sinh(ky), \quad & \quad b_{s,8} = \sinh(kx)\cosh(ky),
\end{align*}
where the numbered indexes are simple values assigned to a particular solution.  However, despite these solutions satisfying \eqref{DiagCartCond1} when $f$ and $g$ vanish, it may then be shown through the use of the compatibility condition $h_{xy}=h_{yx}$ that only a single basis function is not sufficient to satisfy the system as a whole.  Leaving the equations in terms of $b_{s}$, the derivatives of $h$ may be found from \eqref{DiagCartSysEq4},
\begin{align*}
    h_{x} &= -b_{s}\int b_{s}\dd{y}, \\
    h_{y} &= -b_{s}\int b_{s}\dd{x}.
\end{align*}
By taking the second derivative for each of these expressions and setting the difference to zero, we find
\begin{equation}
    \pdv{b_{s}}{x}\int b_{s}\dd{x}-\pdv{b_{s}}{y}\int b_{s}\dd{y} = 0.
    \label{DiagCartCond2}
\end{equation}
It is straightforward to check that indeed a single basis function for $b_{s}$ does not satisfy this relationship.  However, there are sums of two solutions for $b_{s}$ that indeed do satisfy \eqref{DiagCartCond2}, namely
\begin{align}
    b_{s,a} &= b_{s,1}\pm b_{s,3} = \cos(kx)\cos(ky)\pm \sin(kx)\sin(ky) = \cos[k(x\mp y)], \label{DiagCartSol1}  \\
    b_{s,b} &= b_{s,2}\pm b_{s,4} = \cos(kx)\sin(ky)\pm\sin(kx)\cos(kx) = \sin[k(x\pm y)], \\
    b_{s,c} &= b_{s,5}\pm b_{s,7} = \cosh(kx)\cosh(ky)\pm\sinh(kx)\sinh(ky) = \cosh[k(x\pm y)], \\
    b_{s,c} &= b_{s,6}\pm b_{s,8} = \cosh(kx)\sinh(ky)\pm\sinh(kx)\cosh(ky) = \sinh[k(x\pm y)]. \label{DiagCartSol4}
\end{align}
These are the only four linear combinations of two solutions that have been found to satisfy \eqref{DiagCartCond2}.  However, we have also found that there are combinations of the above solutions that also satisfy this condition,
\begin{align*}
    b_{s,\alpha} &= b_{s,a}+b_{s,c}, \\
    b_{s,\beta} &= b_{s,a}+b_{s,d}, \\
    b_{s,\gamma} &= b_{s,b}+b_{s,c}, \\
    b_{s,\delta} &= b_{s,b}+b_{s,d}, \\
    b_{s,\epsilon} &= b_{s,c}+b_{s,d}.
\end{align*}
It is further possible to show that when considering solutions in the form of the above expressions, allowing the two terms to have differing values for $k$ also satisfies \eqref{DiagCartCond2}.  Those solutions of differing $k$'s are identified in Table \ref{Table2}.
\begin{table}
 \caption{Identification of solutions for $b_{s}$ as linear combinations of solutions with differing $k$ values that satisfy \eqref{DiagCartCond2}.}
    \centering
    \begin{tabular}{|c|c|c|c|c|}
    \multicolumn{5}{c}{Non-Maxwellian Solutions} \\
    \hline\hline
     & $b_{s,a}(k_{1})$ & $b_{s,b}(k_{1})$ & $b_{s,c}(k_{1})$ & $b_{s,d}(k_{1})$ \\
    \hline
    $b_{s,a}(k_{2})$ & $\checkmark$ & x & x & x \\
    \hline
    $b_{s,b}(k_{2})$ & x & $\checkmark$ & $\checkmark$ & $\checkmark$ \\
    \hline
    $b_{s,c}(k_{2})$ & x & $\checkmark$ & $\checkmark$ & $\checkmark$ \\
    \hline
    $b_{s,d}(k_{2})$ & x & $\checkmark$ & $\checkmark$ & $\checkmark$ \\
    \hline\hline
    \end{tabular}
    \label{Table2}
\end{table}
These results may be only a subset of a even larger group of non-Maxwellian solutions.  Our investigation into the Cartesian case does not warrant a complete search, as each case listed in Table \ref{Table1} should be considered when suspending the requirement of Maxwellian fields. However, other sub-cases not yet studied are left for future work since they are not aligned with the main thrust of our interests here.

To show that these solutions result in the Poisson bracket vanishing between the invariant and the Hamiltonian we start with $H$, the normalized Hamiltonian in terms of the momenta given by \eqref{eq:normalH}.  If we make use of our magnetic field being of the form \eqref{WantedBField} and let $B=a_{y}$, while $A=0$, we see that the integral over $b_{s}$ with respect to $x$ will give us the potential to use in the Hamiltonian.  Choosing our solution $b_{s,a}$ as in \eqref{DiagCartSol1}, the Hamiltonian becomes
\begin{equation}
    H_{e,n} = p_{x}^{2}+p_{y}^{2}-\frac{2}{k}\sin[k(x-y)]p_{y}+\frac{1}{k^{2}}\sin^{2}[k(x-y)].
    \label{DiagCartHamEx}
\end{equation}
Next, using our system of equations for the coefficients of the invariant, it is a simple matter to find $k_{i}$ and $h$ as
\begin{align*}
    k_{0} &= \frac{1}{k}\sin[k(x-y)], \\
    k_{1} &= \frac{1}{k}\sin[k(x-y)], \\
    h &= -\frac{1}{2k^{2}}\cos^{2}[k(x-y)].
\end{align*}
We may then substitute these results into the invariant to give
\begin{equation*}
    I = \dot{x}\dot{y}+\frac{1}{k}\sin[k(x-y)]\qty(\dot{x}+\dot{y})-\frac{1}{2k^{2}}\cos^{2}[k(x-y)].
\end{equation*}
However, this must also be expressed in terms of the canonical dynamical variables.  Since we are considering a system with unit mass, the canonical momentum is just the linear combination between the velocity and the potential, which we have already specified to only have a $y$-component.  Making the appropriate substitutions then results in
\begin{equation}
    I = p_{x}p_{y}+\frac{1}{k}\sin[k(x-y)]p_{y}+\frac{1}{2k^{2}}\cos^{2}[k(x-y)]-\frac{1}{k^{2}}.
    \label{DiagCartInvExSol}
\end{equation}
With \eqref{DiagCartHamEx} and \eqref{DiagCartInvExSol} in these forms, it is then a straightforward to see that $\comm{H_{e,n}}{I}=0$ holds. We believe that this and similar related results are new in themselves, enlarging the known set of nonlinear, completely integrable Hamiltonians, albeit non-Maxwellian.

\subsection{Cartesian Case}
We return to our main investigation, that is to each non-degenerate case in Table \ref{Table1}, while enforcing the harmonic condition, i.e. the search for physical, nonlinear, integrable magnetic fields.  Starting with the Cartesian case, we recall the general form of the invariant in \eqref{QuadInvariantGeneral},
\begin{equation*}
I = a(x\dot{y}-y\dot{x})^{2}+(x\dot{y}-y\dot{x})(b\dot{x}+d\dot{y})+c\dot{x}^{2}+f\dot{y}^{2}+e\dot{x}\dot{y}+k_{0}\dot{x}+k_{1}\dot{y}+h,
\end{equation*}
and assume that $a=b=d=0$, while $c\neq f$, and $e/(c-f)\neq \pm i$.  Our first integrability-preserving transformation that we may use to reduce the form of the invariant is that of a rotation in the coordinates and velocities \cite{HietarintaDirectSearch},
\begin{align*}
\dot{x}^{\prime} &= \dot{x}\cos\phi+\dot{y}\sin\phi, \\
\dot{y}^{\prime} &= -\dot{x}\sin\phi+\dot{y}\cos\phi,
\end{align*}
such that $e$ may be set to zero by choosing a value for $\phi$ at which the coefficient of the resulting $\dot{x}\dot{y}$ term vanishes, leaving only the terms that are purely dependent upon either $\dot{x}$ or $\dot{y}$.  This may then be further reduced by taking a linear combination with an multiple of the Hamiltonian, $\lambda H$, such that we may set $f\rightarrow0$ with $\lambda=-2f$.  The invariant then only has a quadratic dependence upon $\dot{x}$, and so we scale the resulting coefficient of $c-f$, giving a much reduced form of the invariant,
\begin{equation}
    I = \frac{1}{2}\dot{x}^{2}+k_{0}\dot{x}+k_{1}\dot{y}+h.
\end{equation}
Our system of equations then reduces to
\begin{align}
    g_{0} = \frac{1}{2}, \quad & \quad g_{1} = 0, \\
    g_{2} = 0, \quad & \quad k_{0,x} = 0, \\
    k_{1,y} = 0, \quad & \quad k_{0,y}+k_{1,x} = -b_{s}, \\
    h_{x} = k_{1}b_{s}, \quad & \quad h_{y} = -k_{0}b_{s}.
\end{align}
Since $k_{0x}=0$, we may write that $k_{0}$ is purely some function of $y$ and likewise for $k_{1}$ as purely some function of $x$:
\begin{align}
k_{0} &= -g_{y}(y), \\
k_{1} &= -f_{x}(x).
\end{align}
Taking the respective derivatives of $k_{0}$ and $k_{1}$, we find that
\begin{equation}
    b_{s} = f_{xx}(x)+g_{yy}(y).
\end{equation}
This equation is separable in Cartesian coordinates, leading to linear solutions in $x$ and $y$, and therefore nonlinear Maxwellian fields cannot be found for the Cartesian case.

\subsection{Parabolic Case}
Recalling the general form of the invariant in (\ref{QuadInvariantGeneral}), we start by assuming $a=0$ and $b\neq 0$, such that
\begin{equation*}
I = \qty(x\dot{y}-y\dot{x})\qty(b\dot{x}+d\dot{y})+c\dot{x}^{2}+f\dot{y}^{2}+e\dot{x}\dot{y}+k_{0}\dot{x}+k_{1}\dot{y}+h.
\end{equation*}
We begin simplifying this by performing a rotation, such that $d$ goes to zero.  This is then followed by a linear combination with the Hamiltonian, further reducing the invariant by bringing $f\rightarrow0$.  Additionally, we make a translation of $x\mapsto x-e/b$ and $y\mapsto y-c/b$, which further sets $c$ and $e$ to zero.  Finally, we conduct a scaling to set $b=1$, such that the invariant is of the form
\begin{equation}
I = x\,\dot{x}\dot{y}-y\,\dot{x}^{2}+k_{0}\dot{x}+k_{1}\dot{y}+h.
\end{equation}
With this setup, our conditions for the $g_{i}$'s, $k_{i}$'s, and $h$ become
\begin{align}
g_{0} = -y, \quad & \quad g_{1} = x, \\ 
g_{2} = 0, \quad & \quad k_{0,x} = xb_{s}, \\
k_{1,y} = -xb_{s}, \quad & \quad k_{0,y}+k_{1,x} = 2y b_{s}, \\
h_{x} = k_{1}b_{s}, \quad & \quad h_{y} = -k_{0}b_{s}.
\end{align}
Conducting the appropriate integrals with respect to $x$ and $y$ over $xb_{s}$ gives
\begin{align}
k_{0} &= g(y) +\sum_{k=0}\sum_{j=0}\frac{a_{k,j}}{k+2}x^{k+2}y^{j}, \\
k_{1} &= f(x) -\sum_{k=0}\sum_{j=0}\frac{a_{k,j}}{j+1}x^{k+1}y^{j+1},
\end{align}
with $f,g$ arbitrary. We may then make use of the condition $k_{0,y}+k_{1,x}=2yb_{s}$ by taking the appropriate derivatives of the $k_{i}$'s above to find
\begin{equation}
    g_{y}(y)+f_{x}(x) +\sum_{k=0}\sum_{j=1}\frac{j\,a_{k,j}}{k+2}x^{k+2}y^{j-1}-\sum_{k=0}\sum_{j=0}\frac{k+1}{j+1}a_{k,j}x^{k}y^{j+1} = \sum_{k=0}\sum_{j=0}2\,a_{k,j}x^{k}y^{j+1}.
\end{equation}
With the likeness in powers of $x^{k}y^{j+1}$, we can pull the term of $2yb_{s}$ to the left-hand side,
\begin{equation}
g_{y}(y)+f_{x}(x)+\sum_{k=0}\sum_{j=1}\frac{j\,a_{k,j}}{k+2}x^{k+2}y^{j-1}-\sum_{k=0}\sum_{j=0}\frac{k+2j+3}{j+1}a_{k,j}x^{k}y^{j+1} = 0,
\end{equation}
with which we then expand some of the summations, such that we can find terms dependent only upon $x$, $y$, or a mix between them:
\begin{equation}
\begin{split}
0 =& f_{x}(x)+g_{y}(y)+\sum_{k=0}\frac{a_{k,1}}{k+2}x^{k+2}+\sum_{j=0}\frac{2j+3}{j+1}a_{0,j}y^{j+1}-\sum_{j=0}\frac{2j+4}{j+1}a_{1,j}\,xy^{j+1} \\
& +\sum_{k=2}\sum_{j=0}\qty[\frac{j+2}{k}a_{k-2,j+2}-\frac{k+2j+3}{j+1}a_{k,j}]x^{k}y^{j+1}.
\end{split}
\label{ParabolicCondition6Result}
\end{equation}
If we define $\hat{c}$ for the overlap point at which $f_{x}(x)=-g_{y}(y)$, we find expressions for $f(x)$ and $g(y)$, while the terms mixed in $x$ and $y$ give a recursive relationship and a constraint on the coefficients.
\begin{align}
f(x) &= c_{1}-\hat{c}x-\sum_{k=0}\frac{a_{k,1}}{\qty(k+2)\qty(k+3)}x^{k+3}, \\
g(y) &= c_{2}+\hat{c}y-\sum_{j=0}\frac{2j+3}{\qty(j+1)\qty(j+2)}a_{0,j}y^{j+2}, \\
0 &= \qty(\frac{j+2}{k})a_{k-2,j+2}-\qty(\frac{k+2j+3}{j+1})a_{k,j} \quad \text{for all} \quad k\geq 2, \text{ }j\geq 0, \label{ParabolicRecursiveRelation} \\
0 &= a_{1,j}\quad \text{ for all } \quad j\geq 0.
\end{align}
This final restriction can immediately be seen, as (\ref{ParabolicCondition6Result}) only has a single term with the coordinates of the form $xy^{j+1}$.  All other terms are either higher order in $x$, or purely dependent upon either $x$ or $y$.  This forces the whole term to evaluate to zero, which can only be done by setting the coefficients $a_{1,j}=0$.  However, this leads to a further restriction in that all $a_{k,j}=0$ for all $k$ odd, which can be seen from the recursive relation (\ref{ParabolicRecursiveRelation}).  The only possibility for finding a solution for $b_{s}$ and for $I$, then, is if some $a_{k,j}$'s are non-zero when $k$ is even.

If we now consider the harmonic condition, \eqref{HarmonicCondition}, we see that \eqref{ParabolicRecursiveRelation} may be written as
\begin{equation}
    0 = b_{k,j}\qty(\frac{j+2}{k\qty(k-2)!(j+2)!}+\frac{k+2j+3}{(j+1)k!j!}),
\end{equation}
for all $k\geq 2$ and $j\geq 0$.  This condition then enforces that the $b_{k,j}$ in this range evaluate to zero, and the only non-zero constants are $b_{0,0}$ and $b_{0,1}$, leaving the magnetic field in the form
\begin{equation}
    b_{s} = b_{0,0}+b_{0,1}y,
\end{equation}
that is a linear solution.

\subsection{Spherical Case}
We now consider those cases of the invariant in (\ref{QuadInvariantGeneral}) where $a\neq 0$.  With the general form of the invariant, we may then conduct a translation such that $x\to x-d/2a$ and $y\to y+b/2a$, which brings the invariant to
\begin{equation}
I = a\qty(x\dot{y}-y\dot{x})^{2}+\qty(c-\frac{b^{2}}{4a})\dot{x}^{2}+\qty(f-\frac{d^{2}}{4a})\dot{y}^{2}+\qty(e-\frac{bd}{2a})\dot{x}\dot{y}+k_{0}\dot{x}+k_{1}\dot{y}+h.
\label{InvariantGeneral-a=/=0}
\end{equation}
This will be the form of $I$ that will serve as the starting point for the remaining cases.  Additionally, we introduce three constants,
\begin{align}
\lambda_{1} &= \frac{bd}{2a}-e, \\
\lambda_{2} &= \frac{b^{2}-d^{2}}{4a}-c+f, \\
\sigma &= \sqrt{\lambda_{1}^{2}+\lambda_{2}^{2}}.
\end{align}

We recover what is considered the spherical case when setting $\lambda_{1}=\lambda_{2}=0$, such that the coefficient of the $\dot{x}\dot{y}$ term goes to zero, while the coefficient of $\dot{x}^{2}$ and $\dot{y}^{2}$ become equal.  This can then be followed by taking a linear combination with the Hamiltonian, reducing the invariant to
\begin{equation}
I = a\qty(x\dot{y}-y\dot{x})^{2}+k_{0}\dot{x}+k_{1}\dot{y}+h,
\label{SphericalCaseInvariant}
\end{equation}
with the conditions
\begin{align}
g_{0} = y^{2}, \quad & \quad g_{1} = -2xy, \\ 
g_{2} = x^{2}, \quad & \quad k_{0,x} = -2xy b_{s}, \\
k_{1,y} = 2xy b_{s}, \quad & \quad k_{0,y}+k_{1,x} = 2b_{s}\qty(x^{2}-y^{2}), \\
h_{x} = k_{1}b_{s}, \quad & \quad h_{y} = -k_{0}b_{s},
\end{align}
where $a$ has been scaled to 1.

In order to find possible solutions, we again start by searching for the form of the $k_{i}$'s, while utilizing the power series form of $b_{s}$.  With the definitions of $k_{0,x}$ and $k_{1,y}$, we have
\begin{align}
k_{0} &= g(y)-\sum_{k=0}\sum_{j=0}\frac{2\,a_{k,j}}{k+2}x^{k+2}y^{j+1}, \\
k_{1} &= f(x)+\sum_{k=0}\sum_{j=0}\frac{2\,a_{k,j}}{j+2}x^{k+1}y^{j+2}.
\end{align}
Putting these two results together into $k_{0,y}+k_{1,x} = 2b_{s}\qty(x^{2}-y^{2})$, gives
\begin{equation}
0 = f_{x}(x)+g_{y}(y)+\sum_{k=0}\sum_{j=2}\frac{k+j+1}{j}a_{k,j-2}x^{k}y^{j}-\sum_{k=2}\sum_{j=0}\frac{k+j+1}{k}a_{k-2,j}x^{k}y^{j}.
\end{equation}
The summations can then expanded out to $k=j=2$ in order to find the forms of $f(x)$ and $g(y)$, while also providing the following three conditions,
\begin{align}
0 &= \frac{j+4}{j+2}a_{1,j}\quad \text{for all}\quad j\geq 0, \\
0 &= \frac{k+4}{k+2}a_{k,1} \quad \text{for all} \quad k\geq 0, \\
0 &= j\,a_{k-2,j}-k\,a_{k,j-2}, \quad \text{for all} \quad k,j\geq 2,
\end{align}
where the first two were found through an adjustment of indices.  We immediately see that $a_{1,j}=0$ for all $j$ and $a_{k,1}=0$ for all $k$.  Using the recursive relationship, we find $a_{k,j}=0$ for all $k$ odd and $j$ even, and vice versa.  Additionally, we make use of the Taylor series form of the coefficients, $a_{k,j}=b_{k,j}/(k!j!)$ in conjunction with \eqref{HarmonicCondition} such that
\begin{equation}
\begin{split}
0 &= \frac{j\,b_{k-2,j}}{(k-2)!j!}-\frac{k\,b_{k,j-2}}{k!(j-2)!} \\
&= b_{k-2,j}\qty(\frac{j}{(k-2)!j!}+\frac{k}{k!(j-2)!}) \\
&= b_{k,j-2}\qty(\frac{j}{(k-2)!j!}+\frac{k}{k!(j-2)!}),
\end{split}
\end{equation}
which is again true for all $k,j\geq 2$, thereby imposing that $b_{2,0}=b_{0,2}=a_{2,0}=a_{0,2}=0$ and all higher order coefficients for $k$ and $j$ even.  Therefore, the only possible coefficient that is non-zero is $a_{0,0}$, and there are no nonlinear solutions in the spherical case.

\subsection{Elliptic Case}
From the same reduced form of $I$ before the rotation in the preceding section, \eqref{InvariantGeneral-a=/=0}, this time we perform rotations, linear combinations with $H$, and scalings to recover
\begin{equation}
I = \qty(x\dot{y}-y\dot{x})^{2}+c\qty(\dot{x}^{2}-\dot{y}^{2})+k_{0}\dot{x}+k_{1}\dot{y}+h,
\end{equation}
with $c=\omega\sigma$, where $\omega$ is a non-zero scaling constant.  This form of $I$ then induces our system of equations:
\begin{align}
g_{0} = y^{2}+c, \quad & \quad g_{1} = -2xy, \\ 
g_{2} = x^{2}-c, \quad & \quad k_{0,x} = -2xy b_{s}, \\
k_{1,y} = 2xy b_{s}, \quad & \quad k_{0,y}+k_{1,x} = 2b_{s}\qty(x^{2}-y^{2}-2c), \\
h_{x} = k_{1}b_{s}, \quad & \quad h_{y} = -k_{0}b_{s}, \label{EllipticCase-hxAndhy}
\end{align}
which again allows us to pursue possible solutions to $b_{s}$ and $I$ with the use of the power series representation of $b_{s}$.

As with previous cases, we begin with the conditions on $k_{0,x}$ and $k_{1,y}$ to recover
\begin{align}
k_{0} &= g(y)-\sum_{k=0}\sum_{j=0}\frac{2\,a_{k,j}}{k+2}x^{k+2}y^{j+1}, \\
k_{1} &= f(x)+\sum_{k=0}\sum_{j=0}\frac{2\,a_{k,j}}{j+2}x^{k+1}y^{j+2}.
\end{align}
We then take the appropriate derivatives, and set their linear combination equal to $2b_{s}(x^{2}-y^{2}-2c)$.  This is then followed by a slight adjustment of indices, yielding
\begin{equation}
\begin{split}
0 =& f_{x}(x)+g_{y}(y)+\sum_{k=0}\sum_{j=2}\frac{2\qty(k+j+1)}{j}a_{k,j-2}x^{k}y^{j}-\sum_{k=2}\sum_{j=0}\frac{2\qty(k+j+1)}{k}a_{k-2,j}x^{k}y^{j} \\
& +\sum_{k=0}\sum_{j=0}4c\,a_{k,j}x^{k}y^{j}.
\end{split}
\label{QuadInvEllitpicalCondition6-Full}
\end{equation}
Conducting the summations of $k$ and $j$ such that all summations start from $k=2$ and $j=2$ produces the terms necessary to find the functions $f(x)$ and $g(y)$, along with three mixed terms that allow us to impose additional constraints on the system.
\begin{align}
f(x) &= c_{1}-\qty(2c\,a_{0,0}+\hat{c})x-c\,a_{1,0}x^{2}+\sum_{k=0}\qty[\frac{a_{k,0}}{k+2}-\frac{2c\,a_{k+2,0}}{k+3}]x^{k+3}, \label{QuadInvElliptical-f(x)} \\[2ex]
g(y) &= c_{2}+\hat{c}y-c\,a_{0,1}y^{2}-\sum_{j=0}\qty[\frac{a_{0,j}}{j+2}+\frac{2c\,a_{0,j+2}}{j+3}]y^{j+3}, \label{QuadInvEllitpical-g(y)} \\[2ex]
\begin{split}
0 = 2&c\,a_{1,1}xy+\sum_{j=2}\qty[2c\,a_{1,j}+\frac{j+2}{j}a_{1,j-2}]xy^{j}+\sum_{k=2}\qty[2c\,a_{k,1}-\frac{k+2}{k}a_{k-2,1}]x^{k}y \\
& +\sum_{k=2}\sum_{j=2}\qty[\frac{k+j+1}{j}a_{k,j-2}-\frac{k+j+1}{k}a_{k-2,j}+2c\,a_{k,j}]x^{k}y^{j},
\end{split}
\end{align}
where $f(x)$ and $g(y)$ have been recovered by integrating the recovered forms of $f_{x}(x)$ and $g_{y}(y)$ in (\ref{QuadInvEllitpicalCondition6-Full}).  We also notice that all of the terms mixed in $x$ and $y$ must independently go to zero, implying the following additional conditions.
\begin{align}
    0 &= 2c\,a_{1,1}, \\
    0 &= 2c\,a_{k,1}-\frac{k+2}{k}a_{k-2,1} \quad \text{for all}\quad k\geq 2, \label{QuadInvEllipticalRecursive1} \\
    0 &= 2c\,a_{1,j}+\frac{j+2}{j}a_{1,j-2} \quad \text{for all}\quad j\geq 2, \label{QuadInvEllipticalRecursive2} \\
    0 &= \frac{k+j+1}{j}a_{k,j-2}-\frac{k+j+1}{k}a_{k-2,j}+2c\,a_{k,j} \quad \text{for all} \quad k,j\geq 2. \label{QuadInvEllipticalRecursive3}
\end{align}
Since $c$ is non-zero, these four additional constraints require that the coefficient $a_{1,1}=0$.  In conjunction with the following three conditions then, this enforces that
\begin{equation}
    0 = a_{k,j}\text{ for all }k\text{ odd, }j\text{ odd}.
\end{equation}

We next move to the calculation of the two derivatives of $h$ in \eqref{EllipticCase-hxAndhy} by making use of (\ref{QuadInvElliptical-f(x)}) and (\ref{QuadInvEllitpical-g(y)}) in $k_{1}$ and $k_{0}$, respectively, and multiplying each with the power series of $b_{s}$.  We find from $h_{x}=k_{1}b_{s}$ that
\begin{equation}
\begin{split}
h_{x} &= \sum_{r=0}\sum_{s=0}c_{1}\,a_{r,s}x^{r}y^{s}-\sum_{r=1}\sum_{s=0}\qty(2c\,a_{0,0}+\hat{c})a_{r-1,s}x^{r}y^{s}-\sum_{r=2}\sum_{s=0}c\,a_{1,0}a_{r-2,s}x^{r}y^{s} \\
& +\sum_{r=3}\sum_{s=0}\qty(\sum_{k+i=r-3}\frac{a_{k,s}a_{i,0}}{i+2})x^{r}y^{s}-\sum_{r=3}\sum_{s=0}\qty(\sum_{k+i=r-3}\frac{2c\,a_{i+2,0}a_{k,s}}{i+3})x^{r}y^{s} \\
& +\sum_{r=1}\sum_{s=2}\qty(\sum_{k+i=r-1}\sum_{j+l=s-2}\frac{2\,a_{k,j}a_{i,l}}{l+2})x^{r}y^{s},
\end{split}
\end{equation}
while $h_{y}=-k_{0}b_{s}$ gives
\begin{equation}
\begin{split}
h_{y} &= -\sum_{r=0}\sum_{s=0}c_{2}\,a_{r,s}x^{r}y^{s}-\sum_{r=0}\sum_{s=1}\hat{c}\,a_{r,s-1}x^{r}y^{s}+\sum_{r=0}\sum_{s=2}c\,a_{0,1}a_{r,s-2}x^{r}y^{s} \\
& +\sum_{r=0}\sum_{s=3}\qty(\sum_{j+l=s-3}\frac{a_{r,j}a_{0,l}}{l+2})x^{r}y^{s}+\sum_{r=0}\sum_{s=3}\qty(\sum_{j+l=s-3}\frac{2c\,a_{r,j}a_{0,l+2}}{l+3})x^{r}y^{s} \\
& +\sum_{r=2}\sum_{s=1}\qty(\sum_{k+i=r-2}\sum_{j+l=s-1}\frac{2\,a_{k,j}a_{i,l}}{i+2})x^{r}y^{s}.
\end{split}
\end{equation}
We may then integrate these two results and expand the summations symmetrically so that we may find the two expressions for $h$ below.  Starting with the integration of $h_{x}$, we obtain
\begin{equation}
\begin{split}
h &=  G(y)+c_{1}\,a_{0,0}x+\qty[c_{1}\,a_{0,1}-\qty(2c\,a_{0,0}+\hat{c})a_{0,0}]\frac{x^{2}}{2}+c_{1}\,a_{0,1}xy-\qty(2c\,a_{0,0}+\hat{c})a_{0,1}\frac{x^{2}y}{2} \\
& +\qty[c_{1}\,a_{2,0}-\qty(3c\,a_{0,0}+\hat{c})a_{1,0}]\frac{x^{3}}{3}+c_{1}\,a_{0,2}xy^{2}+\qty[c_{1}a_{2,1}-c\,a_{1,0}a_{0,1}]\frac{x^{3}y}{3} \\
& +\qty[c_{1}a_{1,2}-\qty(2c\,a_{0,0}+\hat{c})a_{0,2}+a_{0,0}^{2}]\frac{x^{2}y^{2}}{2} \\
& +\qty[c_{1}\,a_{2,2}-\qty(2c\,a_{0,0}+\hat{c})a_{1,2}-c\,a_{1,0}a_{0,2}+2\,a_{1,0}a_{0,0}]\frac{x^{3}y^{2}}{3} \\
\\
& +\sum_{s=3}c_{1}\,a_{0,s}xy^{s}+\sum_{s=3}\qty[c_{1}a_{1,s}-\qty(2c\,a_{0,0}+\hat{c})a_{0,s}+\sum_{j+l=s-2}\frac{2\,a_{0,j}a_{0,l}}{l+2}]\frac{x^{2}y^{s}}{2} \\
& +\sum_{s=3}\qty[c_{1}\,a_{2,s}-\qty(2c\,a_{0,0}+\hat{c})a_{1,s}-c\,a_{1,0}a_{0,s}+\sum_{j+l=s-2}\frac{2\,a_{1,j}a_{0,l}+2\,a_{0,j}a_{1,l}}{l+2}]\frac{x^{3}y^{s}}{3} \\
& +\sum_{r=4}\qty[c_{1}\,a_{r-1,0}-\qty(2c\,a_{0,0}+\hat{c})a_{r-2,0}-c\,a_{1,0}a_{r-3,0}+\sum_{k+i=r-4}a_{k,0}\qty(\frac{a_{i,0}}{i+2}-\frac{2c\,a_{i+2,0}}{i+3})]\frac{x^{r}}{r} \\
& +\sum_{r=4}\qty[c_{1}\,a_{r-1,1}-\qty(2c\,a_{0,0}+\hat{c})a_{r-2,1}-c\,a_{1,0}a_{r-3,1}+\sum_{k+i=r-4}a_{k,1}\qty(\frac{a_{i,0}}{i+2}-\frac{2c\,a_{i+2,0}}{i+3})]\frac{x^{r}y}{r} \\
& +\sum_{r=4}\qty[c_{1}\,a_{r-1,2}-\qty(2c\,a_{0,0}+\hat{c})a_{r-2,2}-c\,a_{1,0}a_{r-3,2}+\sum_{k+i=r-4}a_{k,2}\qty(\frac{a_{i,0}}{i+2}-\frac{2c\,a_{i+2,0}}{i+3})]\frac{x^{r}y^{2}}{r} \\
& +\sum_{r=4}\qty(\sum_{k+i=r-2}a_{k,0}a_{i,0})\frac{x^{r}y^{2}}{r}+\sum_{r=4}\sum_{s=3}\qty(\sum_{k+i=r-2}\sum_{j+l=s-2}\frac{2\,a_{k,j}a_{i,l}}{l+2})\frac{x^{r}y^{s}}{r} \\
& +\sum_{r=4}\sum_{s=3}\qty[c_{1}\,a_{r-1,s}-\qty(2c\,a_{0,0}+\hat{c})a_{r-2,s}-c\,a_{1,0}a_{r-3,s}+\sum_{k+i=r-4}a_{k,s}\qty(\frac{a_{i,0}}{i+2}-\frac{2c\,a_{i+2,0}}{i+3})]\frac{x^{r}y^{s}}{r}.
\end{split}
\label{QuadInvElliptical-hFromhx}
\end{equation}
At the same time we integrate $h_{y}$,
\begin{equation}
\begin{split}
h &= F(x)-c_{2}\,a_{0,0}y-c_{2}\,a_{1,0}xy-\qty[c_{2}\,a_{0,1}+\hat{c}\,a_{0,0}]\frac{y^{2}}{2}-\hat{c}\,a_{1,0}\frac{xy^{2}}{2}-c_{2}\,a_{2,0}x^{2}y \\
& -\qty[c_{2}\,a_{2,1}+\hat{c}\,a_{2,0}-a_{0,0}^{2}]\frac{x^{2}y^{2}}{2}-\qty[c_{2}\,a_{2,2}+\hat{c}\,a_{2,1}-c\,a_{0,1}a_{2,0}-2\,a_{0,1}a_{0,0}]\frac{x^{2}y^{3}}{3} \\
& -\qty[c_{2}\,a_{1,2}-c\,a_{0,1}a_{1,0}]\frac{xy^{3}}{3}-\qty[c_{2}\,a_{0,2}+\hat{c}\,a_{0,1}-c\,a_{0,1}a_{0,0}]\frac{y^{3}}{3} \\
& -\sum_{r=3}c_{2}\,a_{r,0}x^{r}y-\sum_{r=3}\qty[c_{2}\,a_{r,1}+\hat{c}\,a_{r,0}-\sum_{k+i=r-2}\frac{2\,a_{k,0}a_{i,0}}{i+2}]\frac{x^{r}y^{2}}{2} \\
& -\sum_{r=3}\qty[c_{2}\,a_{r,2}+\hat{c}\,a_{r,1}-c\,a_{0,1}a_{r,0}-\sum_{k+i=r-2}\frac{2\qty(a_{k,1}a_{i,0}+a_{k,0}a_{i,1})}{i+2}]\frac{x^{r}y^{3}}{3} \\
& -\sum_{s=4}\qty[c_{2}\,a_{0,s-1}+\hat{c}\,a_{0,s-2}-c\,a_{0,1}a_{0,s-3}-\sum_{j+l=s-4}a_{0,j}\qty(\frac{a_{0,l}}{l+2}+\frac{2c\,a_{0,l+2}}{l+3})]\frac{y^{s}}{s} \\
& -\sum_{s=4}\qty[c_{2}\,a_{1,s-1}+\hat{c}\,a_{1,s-2}-c\,a_{0,1}a_{1,s-3}-\sum_{j+l=s-4}a_{1,j}\qty(\frac{a_{0,l}}{l+2}+\frac{2c\,a_{0,l+2}}{l+3})]\frac{xy^{s}}{s} \\
& -\sum_{s=4}\qty[c_{2}\,a_{2,s-1}+\hat{c}\,a_{2,s-2}-c\,a_{0,1}a_{2,s-3}-\sum_{j+l=s-4}a_{2,j}\qty(\frac{a_{0,l}}{l+2}+\frac{2c\,a_{0,l+2}}{l+3})]\frac{x^{2}y^{s}}{s}  \\
& +\sum_{s=4}\qty[\sum_{j+l=s-2}a_{0,j}a_{0,l}]\frac{x^{2}y^{s}}{s}+\sum_{r=3}\sum_{s=4}\qty(\sum_{k+i=r-2}\sum_{j+l=s-2}\frac{2\,a_{k,j}a_{i,l}}{i+2})\frac{x^{r}y^{s}}{s} \\
& -\sum_{r=3}\sum_{s=4}\qty[c_{2}\,a_{r,s-1}+\hat{c}\,a_{r,s-2}-c\,a_{0,1}a_{r,s-3}-\sum_{j+l=s-4}a_{r,j}\qty(\frac{a_{0,l}}{l+2}+\frac{2c\,a_{0,l+2}}{l+3})]\frac{x^{r}y^{s}}{s}.
\end{split}
\label{QuadInvElliptical-hFromhy}
\end{equation}
Since these two expressions for $h$ should be the same, the two functions $F(x)$ and $G(y)$ become
\begin{align}
\begin{split}
F(x) &= c_{1}\,a_{0,0}x+\qty[c_{1}\,a_{1,0}-\qty(2c\,a_{0,0}+\hat{c})a_{0,0}]\frac{x^{2}}{2}+\qty[c_{1}\,a_{2,0}-\qty(3c\,a_{0,0}+\hat{c})a_{1,0}]\frac{x^{3}}{3} \\
& +\sum_{r=4}\qty[c_{1}\,a_{r-1,0}-\qty(2c\,a_{0,0}+\hat{c})a_{r-2,0}-c\,a_{1,0}a_{r-3,0}]\frac{x^{r}}{r} \\
& +\sum_{r=4}\qty[\sum_{k+i=r-4}\qty(\frac{a_{k,0}a_{i,0}}{i+2}-\frac{2c\,a_{k,0}a_{i+2,0}}{i+3})]\frac{x^{r}}{r},
\end{split} \\[3ex]
\begin{split}
G(y) &= -c_{2}\,a_{0,0}y-\qty(c_{2}\,a_{0,1}+\hat{c}\,a_{0,0})\frac{y^{2}}{2}-\qty(c_{2}\,a_{0,2}+\hat{c}\,a_{0,1}-c\,a_{0,1}a_{0,0})\frac{y^{3}}{3} \\
& -\sum_{s=4}\qty[c_{2}\,a_{0,s-1}+\hat{c}\,a_{0,s-2}-c\,a_{0,1}a_{0,s-3}-\sum_{j+l=s-4}\qty(\frac{a_{0,j}a_{0,l}}{l+2}+\frac{2c\,a_{0,j}a_{0,l+2}}{l+3})]\frac{y^{s}}{s}.
\end{split}
\end{align}
Additionally, when comparing the terms that are mixed in $x$ and $y$, we recover a total of 16 constraints on the system.

The possibility for solutions may be further investigated by breaking down the values of the constants $c_{1}$, $c_{2}$, and $\hat{c}$ for when they are all zero, when they are all non-zero, and all possible cases in between.  We consider here the most general case of non-zero $c_{1}$, $c_{2}$, and $\hat{c}$.  We also note that from the harmonic condition, the remaining lowest order coefficients that will determine the remainder of the system are: $b_{0,0}$, $b_{0,1}$, $b_{1,0}$, and $b_{0,2}$. We recall that $a_{k,j}=b_{k,j}/(k!j!)$ and $b_{k,j+2}=-b_{k+2,j}$, such that through this relationship of the $b_{k,j}$'s and the relationships in \eqref{QuadInvEllipticalRecursive1}-\eqref{QuadInvEllipticalRecursive3}, all other coefficients may be found from these four.

We start by considering the relationship from the $xy$ terms in $h$ and see that
\begin{equation}
a_{0,1} = -\frac{c_{2}}{c_{1}}a_{1,0}, \label{QuadInvEllipticalh-Conditionxy}
\end{equation}
which may also be written identically in terms of $b_{1,0}$ and $b_{0,1}$.  This may then be followed up by utilizing the relationship between the $xy^{2}$,
\begin{equation*}
a_{0,2} = -\frac{\hat{c}}{2c_{1}}a_{1,0},
\end{equation*}
or in terms of the $b_{k,j}$ coefficients and utilizing their symmetric property, we can write
\begin{equation}
b_{0,2} = -b_{2,0}=-\frac{\hat{c}}{c_{1}}b_{1,0},
\end{equation}
and so we immediately see that if any one of these coefficients are zero, then there will be no solutions in the elliptical case due to the recursive relationships found earlier as well as the symmetric property of the $b_{k,j}$'s.  It can then be shown that indeed $b_{1,0}$ evaluates to zero by utilizing the relationship from the $x^{3}y$ terms in both forms of $h$ where
\begin{equation}
    c_{1}\,a_{2,1}-c\,a_{1,0}a_{0,1} = -3c_{2}\,a_{3,0}.
\end{equation}
Writing this in terms of the $b_{k,j}$ coefficients and using that $b_{k+1,j-1}=-b_{k-1,j+1}$, such that $b_{3,0}=-b_{1,2}$, we find
\begin{equation}
c_{1}b_{2,1}-c_{2}b_{1,2} = 2cb_{1,0}b_{0,1}.
\end{equation}
Next, we utilize the recursive relationships in \eqref{QuadInvEllipticalRecursive1}-\eqref{QuadInvEllipticalRecursive2} to write this left-hand side in terms of $b_{0,1}$ and $b_{1,0}$, which reduces this to
\begin{equation}
    \frac{c_{1}}{c}b_{0,1}+\frac{c_{2}}{c}b_{1,0} = cb_{1,0}b_{0,1}.
\end{equation}
One can then see that if we were to ensure that $b_{1,0}=b_{1,0}$, we could utilize \eqref{QuadInvEllipticalh-Conditionxy}, which results in the left-hand side evaluating to zero, while the right-hand side evaluates to $(c_{2}c/c_{1})\,b_{1,0}^{2}$.  Since $c$, $c_{1}$, and $c_{2}$ are all taken to be non-zero, this is only true when $b_{1,0}$ is zero.  However, we have seen that $b_{1,0}$ directly relates to $b_{0,1}$, and $b_{2,0}$, and so we find that no solutions exist in the elliptical case.  It should also be mentioned that a null result also comes from the other cases of setting constants to zero; the only influence of setting these constants to zero has on the system is the rate at which the null result is found.

\section{Higher Order Invariants}

The next obvious question is to ask whether there exist invariants of higher degree in the momenta, and whether we can apply these techniques to find them. We will write our general $N$th degree invariant as
\begin{equation}
    I = \sum_{n=0}^N \sum_{i=0}^n f_{n,i} \dot{x}^{n-i} \dot{y}^i,    
\end{equation}
where the $f_{n,i}$ are functions of $x$ and $y$. Once again setting the time derivative to zero yields the following system:
\begin{align}
    \frac{\partial}{\partial x} f_{N,0} &= 0, \\
    \frac{\partial}{\partial x} f_{N,i} + \frac{\partial}{\partial y} f_{N, i-1} &= 0, \quad 1 \leq i \leq N \\
    \frac{\partial}{\partial y} f_{N,N} &= 0, \\
    -f_{n,1} b_s + \frac{\partial}{\partial x}f_{n-1, 0} &= 0, \quad 1 \leq n \leq N \\
    f_{n,n-1} b_s + \frac{\partial}{\partial y} f_{n-1,n-1} &= 0, \quad 1 \leq n \leq N \\
    ((n-i+1)f_{n,i-1} - (i+1)f_{n, i +1})b_s + \frac{\partial}{\partial x} f_{n-1,i} + \frac{\partial}{\partial y}f_{n-1, i-1} &= 0,
\end{align}
where the last is true for $2\leq n\leq N$ and $1\leq i \leq n-1$.
As with the linear and quadratic cases, the first three equations in the system above define our leading coefficients as $N$th degree polynomials in $x$ and $y$. In general the same methods as applied in the linear and quadratic cases become intractable very quickly. However, we notice a general pattern, similar to the lower degree cases, by which the structure of the functions $f_{n,i}$ of degree $N$ are invariant with respect to $N$. As we have seen, these are the exact functions that are the main impediments in finding solutions. The lower degree functions make things intractable quickly, but the structure of the system is highly suggestive of an empty solution set.

To illustrate this more clearly, we have worked out some cases when $N=3$ or $4.$ For this we will drop the double subscripts and adopt the simpler notation as in the linear and quadratic cases. When $N=3$, our invariant is of the form
\begin{equation}
    I = f_0\dot{x}^3 + f_1 \dot{x}^2\dot{y} + f_2 \dot{x}\dot{y}^2 + f_3 \dot{y}^3 + g_0 \dot{x}^2 + g_1 \dot{x}\dot{y} + g_2 \dot{y}^2 + k_0 \dot{x} + k_1 \dot{y} + h,
\end{equation}
and the general system becomes
\begin{align}
f_{0,x} &= 0 \\
f_{3,y} &= 0 \\
f_{0,y} + f_{1,x} &= 0 \\
f_{1,y} + f_{2,x} &= 0 \\
f_{2,y} + f_{3,x} &= 0 \\
-f_{1}b_{s} + g_{0,x} &= 0 \\
3f_{0}b_{s} - 2f_{2}b_{s} + g_{0,y} + g_{1,x} &= 0 \\
2f_{1}b_{s} - 3f_{3}b_{s} + g_{1,y} + g_{2,x} &= 0 \\
f_{2}b_{s} + g_{2,y} &= 0 \\
-g_{1}b_{s} + k_{0,x} &= 0 \\
g_{1}b_{s} + k_{1,y} &= 0 \\
2g_{0}b_{s} - 2g_{2}b_{s} + k_{0,y} + k_{1,x} &= 0 \\
h_x - k_1 b_{s} &= 0 \\
h_y + k_{0} b_{s} &= 0.
\end{align}
Solving for the $f_i$ we obtain cubics in $x$ and $y$
\begin{align}
f_{0} &= a_{3}y^3 + a_{2}y^2 + a_1y + a_{0} \\
f_{1} &= -3a_{3}y^2x - 2a_{2} yx - a_{1} x + d_{2} y^2 + b_{1} y + b_{0} \\
f_{2} &= 3a_{3} x^2y - 2d_{2} xy - d_{1} y + a_{2} x^2 - b_{1} x + c_{0} \\
f_3 &= -a_{3} x^3 + d_{2} x^2 + d_{1} x + d_{0}.
\end{align}
We can then mimic the quadratic case and rewrite our invariant as
\begin{equation}
\begin{split}
I &= a_{3}(x\dot{y} - y\dot{x})^3 + (a_2\dot{x} + d_2\dot{y})(x\dot{y} - y\dot{x})^2 + (d_1\dot{y}^2 - b_1\dot{x}\dot{y} - a_{1}\dot{x}^2)(x\dot{y} - y \dot{x}) + a_{0}\dot{x}^3 \\ 
&+ b_{0}\dot{x}^2\dot{y} + c_0\dot{x}\dot{y}^2 + d_{0}\dot{y^3} + g_0\dot{x}^2 + g_1\dot{x}\dot{y} + g_{2}\dot{y}^2 + k_{0}\dot{x} + k_{1} \dot{y} + h.
\end{split}
\end{equation}
One major barrier we encounter when $N > 2$ is that the invariants do not organize neatly into orbits as they do in the quadratic case. Without this, we are left to try to mimic some of the orbits in the quadratic case and apply similar methods. When we do so, we find that we must then even further simplify the systems by setting various terms to zero in a rather ad hoc method. Any of the systems we have found this way that were manageable by hand computations were then too tightly constrained to yield a solution. We will briefly summarize the most explored case and the problems that arise. For this we consider an analog of the Cartesian invariant of the form
\begin{equation}
    I = a_0\dot{x}^3 + b_0\dot{x}^2\dot{y} + c_0 \dot{x}\dot{y}^2 + d_0\dot{y}^3 + g_0 \dot{x}^2 + g_1 \dot{x} \dot{y} + g_2 \dot{y}^2 + k_1\dot{x} + k_2 \dot{y} + h.
\end{equation}
This does not simplify neatly by way of transformations or linear combinations with the Hamiltonian. However when we impose some more restrictions, we do manage to find invariants, but ones of no physical significance ($b_s$ is forced to either be constant or linear in a single variable). A summary of our work in this case is as follows: $i)$ when $b_{0}=c_{0}=g_{1}=0$, $b_{s}$ is reduced to either being a constant or a function of a single variable, $ii)$ when $b_{0}=c_{0}=d_{0}$ and $a\neq 0$, $b_{s}$ is a linear function of a single variable, $iii)$ for $b_{0}=c_{0}=g_{0}=g_{2}=0$, while $a_{0}$ and $d_{0}$ are nonzero, $b_{s}$ is forced to be invariant in arbitrary translations, which we have seen is unphysical, and $iv)$, when $a_{0}=d_{0}$ and $c_{0}=\pm ib_{0}$, we find that $b_{s}$ must be constant.

We conjecture that due to the issues that arise in even the $N=3$ case, invariants that are of higher order in the momenta do not exist, though a more comprehensive search may be needed.

\section{Conclusion}

Complete integrability in the sense of Liouville has been a topic of interest in Hamiltonian dynamics for a long time. Over the last decade, it became clear that particle accelerators at the intensity frontier might benefit from improved understanding of nonlinear Hamiltonians that are both completely integrable and represent motion of charged particles in external static magnetic fields. Indeed, under certain approximations in the Hamiltonian (paraxial limit) and Maxwell equations (magnetic fields can be obtained from a single non-zero component of the vector potential), completely integrable nonlinear Hamiltonians exist with invariants quadratic in momenta. IOTA is the prime example that fits in this class.

We performed the first systematic search of completely integrable nonlinear Hamiltonians without the approximations described above. The existence of these systems is tied to higher order symmetry principles. We proved that the set of such systems with second invariant quadratic in the momenta is empty, although those systems not constrained by Maxwell's equations do admit quadratic integrals of motion. We have also conjectured the same result for systems with second invariants that are of higher degree in the momenta, based on a few cases worked out explicitly. Proving in general the case of higher order invariants is made difficult by the complexity of the systems of equations to be solved, and the breakdown of symmetry-based arguments. We also showed that finite length solenoids, with their fringe fields, are completely integrable with second invariants linear in the momenta, which is a consequence of basic symmetries according to Noether's theorem.

The empty quadratic solution set is a consequence of the constraints due to the Laplace equation (with or without approximations); without the Laplace constraint completely integrable nonlinear Hamiltonians can be found. Hence, the well-known Courant-Snyder theory of linear (completely integrable) Hamiltonian systems cannot be generalized to the nonlinear case. Therefore, other than the less appealing route of searching for non-autonomous transcendental invariants, approximations in the Hamiltonian need to be made in order to find new integrable systems. Thus, a promising research avenue remains: keeping the paraxial approximation in the Hamiltonian and extending it to 6D phase space in order to accommodate acceleration.

Finally, the set of physically unconstrained solutions seems to show a remarkable richness that is worth pursuing on its own, regardless of their application to beams and accelerators. There is already a vast literature on the topic.

\section{Acknowledgement}

This work was supported in part by the U.S. Department of Energy, Office of High Energy Physics, under Contract No. DE-SC0020064, with Northern Illinois University.

\bibliography{bibliography}

\end{document}